\documentstyle[12pt]{article}
\begin{document}
\thispagestyle{empty}
\setlength{\baselineskip} {4.0ex}
\newcommand{\preprint}[1]{\begin{table}[t]\begin{flushright}
    \begin{large}{#1}\end{large}\end{flushright}\end{table}}
\newcommand{\p}{$\overline{p}$}
\newcommand{\n}{$\overline{n}$}
\newcommand{\N}{$\overline{N}$}
\newcommand{\D}{$\overline{d}$}
\def\ie{{\it i.e.} }
\def\etal{{\it et al.}}
\def\sa{$S/A$ }
\def\pt{$p_T$ }
\def\et{$E_T$ }
\def\fm3{fm$^3$}
\def\fmm3{fm$^{-3}$}
\def\fsi{$FSI$ }
\def\isi{$ISI$ }
\def\min{$M_{in}$ }
\def\sch{Schr$\ddot{o}$dinger }
\def\ham{Hamiltonian }
\def\pr{Phys.\ Rev.\ }
\def\pra{Phys.\ Rev.\ }
\def\prb{Phys.\ Rev.\ }
\def\prc{Phys.\ Rev.\ }
\def\prd{Phys.\ Rev.\ }
\def\prl{Phys.\ Rev.\ Lett.\ }
\def\pl{Phys.\ Lett.\ }
\def\np{Nucl.\ Phys.\ }
\def\ap{Ann.\ Phys.\ (N.Y.) }
\def\prep{Phys.\ Rep.\ }
\def\jp{J.\ Phys.\ }
\def\zp{Z.\ Phys.\ }
\def\mpl{Mod.\ Phys.\ Lett.\ }
\def\rmp{Rev.\ Mod.\ Phys.\ }
\def\sjnp{Sov.\ J.\ Nucl.\ Phys.\ }
\vspace{0.4cm}
\preprint{BGU-94/04\\ hep-ph/9505264}
\title{
$FSI$ Corrections For Near Threshold Meson Production
In Nucleon-Nucleon Collisions}
\vspace{0.4cm}
\author{ A. Moalem $^{}$, L. Razdolskaja $^{}$ and E. Gedalin $^{}$\\
\raggedbottom
        {\small  Department of Physics, Ben Gurion University, 84105
Beer Sheva, Israel}\\
                     }
\date{3 June 1994}
\maketitle
\begin{abstract}
A procedure is proposed which accounts for
final state interaction corrections for near threshold
meson production in nucleon-nucleon scattering. In analogy with the
Watson-Migdal approximation, it is shown that in the limit of extremely
strong final state effects, the amplitude factorizes into a primary
production amplitude and an elastic scattering amplitude describing a
$3 \to 3$ transition.
This amplitude determines the energy dependence of the reaction cross
section near the reaction threshold almost solely.
The approximation proposed satisfies the Fermi-Watson
theorem and the coherence formalism. Application of this procedure
to meson production in nucleon-nucleon scattering shows that, while
corrections due to the meson-nucleon interaction are small for
s-wave pion production,
they are crucial for reproducing the energy dependence of the $\eta$
production cross section.
\end{abstract}
\vspace{2.0 cm}

PACS number(s) : 13.75.Cs, 14.40.Aq, 25.40.Ep

\newpage
\begin{center}
\section{Introduction}
\end{center}

The cross sections for $\pi^0$ and $\eta$ meson production via the
$pp \to pp \pi^0$ and $pp \to pp \eta$ reactions, at energies near their
respective thresholds, exhibit a pronounced energy dependence
which deviates strongly from phase space\cite{meyer,bergd}.
Such deviations, which occur in
other hadronic collisions as well\cite{wilkin},
are most certainly due to final state interactions($FSI$) between the reaction
products.
While $FSI$ corrections due to the short range nucleon-nucleon($NN$)
force between the outgoing protons seem adequate to
reproduce the observed energy dependence\cite{meyer} and the cross section
scale\cite{lee} for pion production, they fail to do so for the $\eta$
production\cite{bergd}. Particularly, several model
calculations\cite{germo,lagt,vett}
have considered the
$NN \to NN \eta$ through a perturbative approach
but none of them reproduced the energy dependence
near threshold, although $FSI$
corrections were made to account for the proton-proton interaction.
It is the purpose of the present work to show that
while the meson-nucleon interaction influences s-wave pion
production only slightly, its effects are crucial
for reproducing the energy dependence of the $pp \to pp \eta$
reaction.

There exists no theory of final state effects in the presence of three
strongly interacting particles and our first objective will be to develop
a procedure that accounts for both the $NN$ and
meson-$N$ forces. To accomplish this task we consider a production process,
$ab \to cde$. A full dynamical theory of such a process would require the
solution of two and three-body scattering problems. Particularly, in the
Faddeev formalism\cite{faddeev} the full transition amplitude is
decomposed as a sum of three terms depending on which of the three final
particle pairs interacts last. The evaluation of these terms, the so called
Faddeev amplitudes, requires the solution of a set of coupled integral
equations. Such a procedure should resolve the problem at hand
but, in view of the scarcity of knowledge about
the $\eta N$ interaction it may turn to be unreliable and rather long and
tedious to apply.
In what follows
we develop an approximation along exactly the lines of the Watson's
approach\cite{watson} for two body reactions, $\ie$, we look for a separation
of the energy dependence due to $FSI$ from those of the primary production
amplitude. Such an approximation may prove useful
and easy to apply in analyzing near
threshold meson production data from $NN$ scattering.

According to the Watson-Migdal $FSI$ theory for two body
processes\cite{watson},
the energy dependence of the full transition amplitude is
governed by rescattering of final state particles.
The full transition amplitude is approximated by,
\begin{equation}
T_{if} \approx M^{(in)}_{if} \ t^{(el)}_{ff}\ \ ,
\label{eq:11}
\end{equation}
where the primary production amplitude,
$M^{(in)}_{if}$, is assumed to be a smooth and slowly varying function of
energy and $t^{(el)}_{ff}$ is taken as the $free$ two-body elastic scattering
amplitude in the exit channel.
Thus the energy dependence of the full matrix element is
determined mostly by  $t^{(el)}_{ff}$.
A decomposition as such of the amplitude into a
primary production amplitude and a subsequent $FSI$ correction term is
meaningful only if $t^{(el)}_{ff}$ is sufficiently large so that \fsi dominate
the reaction amplitude.
This requires that the distortion of the final state wave function
(\ie the deviation
from a simple product of free particle wave functions) or alternatively,
the $sticking\  factor$ (\ie the probability of the particles to
find each other's vicinity) in itself be proportional to $t^{(el)}_{ff}$.
Albeit, the particles spend a great deal of time close
together, as should be the case in  particle production processes
close to threshold.
Otherwise, the distortion factor of the final state (and obviously the full
matrix element) must be calculated by using more precise
methods\cite{watson}.

The complete transition amplitude of a production process can
be decomposed in the form shown in fig. 1, where the first term (the block
$M^{(in)}$) represents the primary production amplitude which includes
all possible inelastic transitions contributing to the process,
while all other terms represent the
contributions from rescatterings in the entrance and exit channels (the
\isi and \fsi blocks).
The third term, for example, corresponds to a process in
which the primary production
takes place as if there were no $FSI$ and only
subsequently is distorted by the Coulomb and nuclear short range interactions
between the reaction products.
At energies near the reaction threshold, where the relative
energies of particle pairs in the final state are sufficiently small
and \fsi corrections are dominating, such a term is expected to be more
important than others.
In analogy to the Watson-Migdal approximation we make the $ansatz$ that in the
limit of weak \isi and strong \fsi, the transition amplitude of a three-body
reaction can be approximated by an expression similar to eqn. 1, with the
\fsi correction term being replaced by the on mass-shell $elastic$
scattering amplitude of
a $3 \to 3$ (three particles in to three particles out) transition.
This amplitude is denoted by $T^{(el)}_{33}$ and,
can be evaluated using the Faddeev formalism\cite{thomas}
or others such as the Weinberg method\cite{weinberg}.
Certainly, it is not a measurable quantity
as such a $3 \to 3$ transition is not easy to realize experimentally.
It is shown in sect. 2 though,
that at energies near
particle production threshold, $T^{(el)}_{33}$ can be estimated from
two-body scattering data
without having to solve the full three-body dynamics.

Within this approximation, the transition amplitude of a three-body reaction
is coherent in terms of the interactions between
the various final particle pairs, just as one anticipates based on general
quantum mechanical considerations\cite{amado}.
Thus the effects due to a given pair's interaction is distributed over the
entire transition amplitude and may influence strongly the energy dependence
of the cross section.
In sect. 3 we apply the procedure developed to
production of mesons in $NN$ scattering and show that although
the $\eta-N$ interaction is rather weak with respect to the $NN$ interaction
it makes remarkably strong contributions.
Discussion of the results and conclusions are given in sect. 4.

\begin{center}
\section{Theoretical Perspective}
\end{center}

\subsection*{2.1 Two-body reactions}
Prior to extending the Watson-Migdal approximation to
three-body reactions let us recall first the more familiar problem of
\fsi corrections in two-body reactions. Assume that the full interaction
between particles separates into $v\ =\ w\ +\ s$, where $w$ and $s$
stand for weak and strong terms respectively. Here $w$ is the
primary interaction, such that if it were zero,
the process in question would not
occur. The remaining part of the interaction $s$ is responsible for
rescatterings in the entrance and exit channels. We further assume that $s$
is the only strong channel energetically allowed.

We would like to find how $FSI$ affects the transition
amplitude in a process in which the primary interaction is relatively weak.
The full transition amplitude in an appropriate eigenchannel is defined as,
\begin{equation}
 T_{if}  = \langle \Psi^{(-)}_{el,f}
|\ \hat{M}^{(in)}_{if}\ | \Psi^{(+)}_{el,i}\rangle \ \ ,
\label{eq:21}
\end{equation}
where $\Psi^{(\pm)}_{el,i(f)}$ are two-particle scattering
wave functions. They correspond to solutions of the \sch equation with
a Hamiltonian
that contains strong $s$-interactions only and satisfy the boundary
conditions that $\Psi^{(+)}_{el,i}$ ($\Psi^{(-)}_{el,f}$) tends asymptotically
to free two-particle wave function $\phi_{oi}$ ($\phi_{of}$),
plus outgoing (ingoing) spherical waves.
These can be written in the form\cite{thomas},
\begin{equation}
\Psi^{(\pm)}_{el,\lambda}(E)  =
\lim_{\epsilon \to 0} \ [\ 1\ +\ G_{02} (E \pm i\epsilon)\
\hat{t}^{(el)}_{\lambda \lambda}]\  \phi_{0\lambda} \ \ , \lambda = i,\ f\ \ .
\label{eq:22}
\end{equation}
Here $G_{02} (E \pm i\epsilon)$ is a free two-body Green's functions, and
$t^{(el)}_{ii}$ and $t^{(el)}_{ff}$ are two-body elastic
scattering amplitude for the entrance and exit channels, respectively.
Substituting eqn. 3 in 2 leads to the full transition amplitude,
\begin{equation}
\hat{T}_{if}  = \hat{M}^{(in)}_{if} +  \hat{t}^{(el)}_{ii} G_{02}
\hat{M}^{(in)}_{if} +
\hat{M}^{(in)}_{if} G_{02} \hat{t}^{(el)}_{ff} +
\hat{t}^{(el)}_{ii} G_{02} \hat{M}^{(in)}_{if} G_{02}
\hat{t}^{(el)}_{ff}\ \ .
\label{eq:210}
\end{equation}
Note that this expression is exact and
contains both \isi and \fsi.
In the Watson-Migdal approximation,
\begin{equation}
 T_{if}  \approx \langle \Psi^{(-)}_{el,f}\ |\ \phi_{0f}\rangle
\langle\ \phi_{0f}\
|\ \hat{M}^{(in)}_{if}\ | \phi_{0i}\rangle \langle\
\phi_{0i} |\Psi^{(+)}_{el,i}\rangle \ \ ,
\label{eq:23}
\end{equation}
where the two overlapping integrals
$\langle\phi_{0i} |\Psi^{(+)}_{el,i}\rangle$
and $\langle \Psi^{(-)}_{el,f}\ |\ \phi_{0f}\rangle$
are on mass-shell $S$ matrix elements which
contain \isi and \fsi corrections, respectively.
To cast the amplitude, eqn. 5, in the form given by the Watson's approximation,
eqn. 1, one has to replace $\Psi^{(+)}_{el,i}$ by $\phi_{0i} $ so that $S_{ii}$
is equal to unity and, in the limit of extremely strong \fsi,
take the on mass-shell $t^{(el)}_{ff}$ instead of
the $S_{ff}$ matrix elements.

It would be instructive to repeat these arguments using a diagrammatic
language. We associate
the weak primary interaction $w$ with the diagram $2a$, and the
strong $s$ interaction with the diagram 2b of figure 2. As usual the
disconnected diagram $2c$ describes noninteracting particles.
It is evident that the amplitude
$T_{if}$ is the sum of all diagrams that can be constructed
from the elementary diagrams $2a$ and $2b$.
To single out the \isi and \fsi blocks and obtain the full matrix element
in a diagrammatic representation similar to the one of fig. 1,
we first sum in the initial and final
channels all diagrams that contain strong interactions only (see fig. 3).
These sums yield the elastic amplitudes
$t^{(el)}_{\lambda \lambda}$ ($\lambda = i,\ f$)
which we identify with
the appropriate \isi and \fsi blocks, respectively.
The sum of all diagrams that start or end with a weak interaction
(diagram $2a$) forms the block $M^{(in)}$ which describes the $primary$
production process.
When rescatterings in both of the initial and final states give sufficiently
large amplitudes, the Migdal-Watson ideology tells us that the
4th term on the rhs of eqn. 4
is dominating. When $FSI$ ($ISI$) effects are small, then the second
(third) term dominates.
Thus the $ISI$ and $FSI$ effects can be separated from the full transition
amplitude by summing diagrams in a specific order
and subsequently derive the Watson-Migdal approximation by
selecting an appropriate dominant term in the limit of extremely
weak \isi and extremely strong \fsi.
We apply this same reasoning to obtain the Watson-Migdal approximation for
three-body reactions.

\subsection*{2.2 Three-body reactions}

Let $T_{23}$ be the transition amplitude of a production process in which
an initial two-body process goes into a final three-body state. For conciseness
assume all particles to be distinguishable and spinless. This restriction
simplifies the mathematics but can be easily removed to treat more
general cases. Furthermore, assume that each pair in the three-body final
state has a given angular momentum so that a unique elastic scattering state
is assigned to each pair simultaneously. This is very often the case
at energies near threshold where,
only one partial wave dominates for each pair namely, $\lambda = 0$.
Following the discussion given above for two-body reactions the interactions
are divided into two categories: (i) elastic interactions leading to
rearrangement of particle (inner) quantum numbers but not to the production of
any additional hadrons, (ii) inelastic interactions which are responsible
for the production (or annihilation) of particles so that particle numbers
in the initial and final states are different. Likewise, we call elastic
those diagrams which involve elastic interactions only and
having equal number of legs to the left hand and to the right hand sides. Those
which begin and end with inelastic interactions and
having different number of legs to the left and right hand sides
are called inelastic.

The transition amplitude $T_{23}$ of the reaction under considerations
is the sum of all diagrams which
start with two legs on the left and end with three legs on the right,
that can be assembled using all possible elastic and inelastic diagrams.
To decompose $T_{23}$ into four terms as depicted in fig. 1, these diagrams
need be organized in a specific order before suming them.
First note that by summing all two-particle elastic diagrams
in the entrance channel and all
three-particle elastic diagrams in the exit channel, one obtains the
$2 \to 2$ and $3 \to 3$ $elastic$ transition amplitude of fig. 3.
These elastic amplitudes which are denoted by $t^{(el)}_{22}$ and
$T^{(el)}_{33}$, contain only
elastic interactions and as such,
they are only parts of the complete $2 \to 2$ and $3 \to 3$ amplitudes.
Although at energies above particle
production threshold, two and three-particle channels
may well be opened in both the initial and final states,
these two amplitudes are calculated by taking into
account only elastic interactions. All inelastic and quasi-elastic
processes are confined into
$M^{(in)}$.
Note also, that the $2 \to 2 $ elastic block contains connected
diagrams only but, the $3 \to 3$ block  contains disconnected diagrams
as well (see fig. 3).

The set of all diagrams contributing to the entire transition
amplitude $T_{23}$,
is now separated into four partial sums each
corresponding to one of the terms in the block diagram of fig. 1 :
(i) the block $M^{(in)}$ is the sum of all inelastic diagrams only,
(ii) the \isi-$M^{(in)}$ term is the sum of all diagrams which start with
a $2 \to 2$ elastic diagram on the left and end with an inelastic diagram
on the right, (iii) the $M^{(in)}$-\fsi term is the sum of all diagrams
which start with an inelastic diagram on the left and end with a $3 \to 3$
elastic diagram to the right and, (iv) the \isi-$M^{(in)}$-\fsi term is
the sum of all diagrams which start with a $2 \to 2$
elastic diagram on the left and end with a $3 \to 3 $
elastic diagram on the right. Obviously, these four
partial sums exhaust the entire set of diagrams that may
contribute to the complete
transition amplitude.

This same presentation of $T_{23}$ as the sum of the terms of
fig. 1 can as well be derived formally.
By definition, the transition amplitude is,
\begin{equation}
T_{23}  =  \langle  \Psi^{(-)}_{el,3} | \hat{M}^{(in)} | \Psi^{(+)}_{el,2}
\rangle \ \ ,
\label{eq:221}
\end{equation}
where as defined in eqn. 3, $\Psi^{(+)}_{el,2}$ stands
for a two-body elastic scattering
wave function and $\Psi^{(-)}_{el,3}$ is a three-particle
elastic scattering
wave function corresponding to the exit channel.
It is a solution of a three-body Schr$\ddot{o}$dinger
equation with a Hamiltonian having only two-body and
three-body elastic interactions, and satisfies the boundary conditions that
$\Psi^{(-)}_{el,3}$ tends asymptotically to a free three-particle wave
function, $\phi_{03}$, plus outgoing spherical waves. In terms of the
elastic scattering amplitudes defined above one has the
expression\cite{thomas},
\begin{eqnarray}
\Psi^{(-)}_{el,3} = \lim_{\epsilon \to 0}\
[\ 1 + G_{03}( E\ +\ i\epsilon )\  \hat{T}^{(el)}_{33}\  ]\ \phi_{03}\ \ .
\label{eq:222}
\end{eqnarray}
Using this with eqn. 3 in eqn. 6 implies that,
\begin{equation}
\hat{T}_{23}  = \hat{M}^{(in)} +  \hat{t}^{(el)}_{22}
\hat{G}_{02} \hat{M}^{(in)} +
\hat{M}^{(in)} \hat{G}_{03} \hat{T}^{(el)}_{33} +
\hat{t}^{(el)}_{22} \hat{G}_{02} \hat{M}^{(in)} \hat{G}_{03}
\hat{T}^{(el)}_{33}\ \ .
\label{eq:223}
\end{equation}
Formally, this expression provides an exact solution of the problem
and has a structure identical in form to that given in eqn. 4, for two-body
reactions. A notable feature of this expression is
that all inelastic transitions
are confined in the block $M^{(in)}$, while all possible $elastic$
rescatterings in the entrance and exit channels are contained
in $t^{(el)}_{22}$ and $T^{(el)}_{33}$.

Let us now consider \fsi corrections $a\ la\ Migdal-Watson$.
In the limit of extremely weak \isi one may replace
$\Psi^{(+)}_{el,2}$ by $\phi_{02}$ in eqn. 6.
This makes the second and fourth terms on the rhs of eqn. 8 disappear.
Furthermore, in the limit of very strong \fsi the third term on the rhs of
eqn. 8 dominates and we may approximate the transition amplitude by,
\begin{equation}
T_{23}  \approx \langle \phi_{02} | \hat{M}^{(in)} | \phi_{03}
\rangle  \ T^{(el)}_{33} \ \ .
\label{eq:224}
\end{equation}
We may argue that the matrix element
$\langle \phi_{02} | \hat{M}^{(in)} | \phi_{03} \rangle $
is a smooth and slowly varying function of the energy and momenta
and hence the energy dependence
of the amplitude, eqn. 9, is determined almost
solely by the elastic amplitude $T^{(el)}_{33}$.
Although in form eqn. 9 resembles the Watson-Migdal approximation
for two-body reactions, eqn. 1, there exists an essential difference between
the two cases which makes eqn. 9 more difficult
to apply. This is because the amplitude $T^{(el)}_{33}$, contrary to
$t^{(el)}_{ff}$, is practically not a measurable quantity.

The two-body initial state,
$\Psi_{el,2}^{(+)}$, can be calculated using
distorted wave approach and the effects of slowly varying
nonvanishing \isi can be
included by replacing $M^{(in)}$ in eqn. 9 with the matrix element
$\langle \phi_{03} | \hat{M}^{(in)} | \Psi^{(+)}_{el,2} \rangle $.
With this modification in mind eqn. 9 is more suitable to apply to data.
Certainly, \isi must be included in a full dynamical description of the
process as they influence the cross section scale through the matrix element
$M^{(in)}$.

\subsection*{2.3 The amplitude for $3 \to 3$ }

In the Faddeev formalism $T^{(el)}_{33}$ is written as a sum of
three Faddeev amplitudes,
\begin{equation}
T^{(el)}_{33} = \sum_{j=1}^{3} T^j\ \ .
\label{eq:8}
\end{equation}
These satisfy the set of coupled integral equations\cite{thomas},
\begin{equation}
T = t + F T  = \sum_{ n=0 }^{\infty} \ (\ FF\ )^n\ [\ t + F t\ ]\ \ ,
\label{eq:8}
\end{equation}
where we have used the notation,
\begin{eqnarray}
T = \left ( \begin{array}{c}
T^1 \\
T^2 \\
T^3 \\ \end{array} \right) \ \ ,
t = \left ( \begin{array}{c}
t_1 \\
t_2 \\
t_3 \\ \end{array} \right) \ \ ,
F = \left ( \begin{array}{clcr}
0 & t_1 & t_1 \\
t_2 & 0 & t_2 \\
t_3 & t_3 & 0 \\ \end{array} \right) G_{03}\ \ ,
\label{eq:2.38}
\end{eqnarray}
and final particles are labeled 1,2 and 3.
Here $t_j$ are two-body scattering amplitudes in the three-body space.
In momentum space
and adopt the convention that ($j, l, k$) are always cyclic,
\begin{equation}
\langle {\bf p}_j,{\bf p}_l,{\bf p}_k\ |\ t_j\ |\ {\bf p}_j',
{\bf p}_l',{\bf p}_k'\rangle
= \langle {\bf p}_l,{\bf p}_k\ |\ \hat{t}_j( E - \epsilon_j)
\ |\ {\bf p}_l',{\bf p}_k'\rangle\ \delta ({\bf p}_j - {\bf p}_j')\ \ ,
\label{eq:8}
\end{equation}
with $\hat{t}_j$ being a solution of the ordinary two-body
Lippmann-Schwinger equation of the energy variable $E - \epsilon_j$.
E is the total three-body energy
and $\epsilon_j$ the energy appropriate to momentum ${\bf p}_j$.
{}From the definitions in eqn. 12,
\begin{equation}
T^j
= t_j \ +\ (T^l + T^k)\ G_{03}\ t_j\ \ ,\ j\neq l\neq k \neq j\ \ ,
\ \ j = 1-3\ \ .
\label{eq:8}
\end{equation}
As pointed by Amado\cite{amado}, the contributions from the various pair
interactions add coherently and the effects of a given pair's interactions
is distributed over the entire amplitude.
{}From eqn. 14, $T^j$ is the sum of all terms
contributing to $T^{(el)}_{33}$ which end with $t_j$ but, it does not include
all contributions of the $l-k$ interaction.
Note that the first term on the rhs represents
the $l-k$ two-body scattering
amplitude with the $j$th particle being a spectator
and as such it corresponds to a disconnected diagram.
In the center of mass of the $l-k$ pair and
for a given partial wave $\lambda$, $t_j$ has the
elastic scattering phase $\delta_{\lambda}(j)$.
The second term in eqn. 14 represents the contributions from
two or more rescatterings and involves completely connected diagrams only.
These play an important role in the coherence formalism\cite{amado} through
interference with similar terms involving the $j-l$ and $j-k$ pair
interactions.
Therefore, in order to account for
the effects of any pair's interaction the entire amplitude must be constructed.
Adding the sum ($T^l + T^k$) to $T^j$ of eqn. 14,
one obtains three equivalent forms of the entire amplitude,
\begin{equation}
T^{(el)}_{33} =
t_j + ( T^l + T^k )\ [ 1 + G_{03} ( E )\  t_j ]\ ,\ j\neq l\neq k \neq j\ \ ,
\ \ j = 1-3\ \ .
\label{eq:8}
\end{equation}
In each of the forms in eqn. 15, there appears the factor
($1 + G_{03}\  t_j$)
with the $t_j$ being half on-shell.
At a given $l-k$ relative momentum (${\bf q}_j$)
in a given $l-k$ partial wave ($\lambda$)
and following the arguments in ref.\cite{amado},
these half-on-shell amplitudes, the factor  ($1 + G_{03}\  t_j$)
and, the entire amplitude have the phase $\delta_{\lambda}$(j).
For this to happen, say in a region where
the expression $(T^l\ + \ T^k)\ G_{03}\ t_j$
may vary rapidly with energy it must have a part which cancels the variation
of $(T^l\ +\ T^k)$ so as to give to their sum the behaviour of $t_j$. This
general observation on three-body amplitudes, which is true for
$T_{23}$ as well as for $T_{33}$, has far reaching consequences on the
energy dependence of
the entire amplitude. Certainly, in order to take this cancelation
into account the
terms corresponding to completely connected diagrams must be included.

Another general constraint on the form of $T_{23}$
is provided by the so called Fermi-Watson theorem.
This requires that the phase of the transition amplitude of the
entire amplitude is determined by the phases and norms of the rescattering
amplitudes in the entrance and exit channel and that in the limit of vanishing
\isi the overall phase of the entire amplitude is equal to
that of the elastic scattering amplitude in the exit channel.
This well established for two-body reactions\cite{joachim}
is easily extended to the case of three-body reactions near threshold
(see Appendix A). Explicitly, for the case under considerations, this theorem
requires that
the phase of $T_{23}$ must be identical to the overall
phase of $T^{(el)}_{33}$.
Both of these constraints are satisfied by the approximation, eqn. 9,
in a natural way.

Turning now to estimate the rescattering amplitude, we note first that
the kernel $F F$ is compact for any complex E and
therefore eqns. 11 has a unique solution.
By solving these equations by iteration , one can calculate the amplitude
$T^{(el)}_{33}$ with the required accuracy.
To first order $T^{(el)}_{33}\  \approx \ [\ t\ +\ Ft\ ]$. It
is the sum of all diagrams shown in fig. 3b.
The first three of these are disconnected and represent
the contributions of the free $t$. As shown by Amado\cite{amado} and
Cahill\cite{cahill} taking the sum of just these three diagrams would
not yield a coherent amplitude and therefore would not be a satisfactory
solution.
The other diagrams of fig. 3b give the first order
correction due to two subsequent elastic scatterings, they are all
completely connected and as indicated above
play an important role in the coherence
mechanism\cite{amado,cahill}.
The next iteration introduces connected diagrams with three and four
rescatterings and so on. Summing the contributions of diagrams up to the
corresponding order leads, in principio, to $T^{(el)}_{33}$ with the
accuracy required.

As a first step we restrict the following discussion
to \fsi corrections due to one and two rescattering diagrams
only (fig. 3b), \ie, we calculate $T^{(el)}_{33}$ to first order only.
Denoting contributions from double scattering diagrams by
$C_{jl}$ it is shown in Appendix B that for s-wave production at
energies close to threshold,
\begin{equation}
C_{jl}  =  M^{(in)} (0,0)\  q_{j}\
t^{(on)}_j (0, \frac {q_{j}^2 }{ 2 \mu_{j}})\
t^{(on)}_l (0, \frac {q_{l}^2 }{ 2 \mu_{l}})\  I_{jl}\ \ .
\label{eq:2.39}
\end{equation}
where $I_{jl}$ are integrals of Kowalski-Noyes\cite{kowalski}
half-shell functions over the appropriate relative momenta of
the interacting pairs. Thus the entire transition amplitude
can be written as,
\begin{equation}
T_{23} \approx M^{(in)}_{23}(0,0)\  Z_{33}\ \ ,
\label{eq:2.39}
\end{equation}
where the \fsi correction factor is given by,

\begin{equation}
Z_{33} = \sum_{j=1}^{3} t^{(on)}_j (0, \frac {q_{j}^2 }{ 2 \mu_{j}})\ +
\sum_{j,l = 1}^{3} q_{j}\ t^{(on)}_j (0, \frac {q_{j}^2 }{ 2 \mu_{j}})\
t^{(on)}_l (0, \frac {q_{l}^2 }{ 2 \mu_{l}})\
I_{jl}\ \ .
\label{eq:2.39}
\end{equation}
A further simplification can be achieved by setting all $I_{jl}$ equal
unity, thus neglecting off shell effects (see Appendix B).
This allows calculating the \fsi factor from two-body scattering data of
the particles involved.
To evaluate $T^{(el)}_{33}$ to second order we need to include
$(\ FF\ )\ [\ t\ +\ Ft\ ]$ terms. The evaluation of such terms is described
in Appendix C, where it is shown that the effect of the (FF) factor is to
scale the first order amplitude $[\ t\ +\ Ft\ ]$ by,
\begin{eqnarray}
\langle \ FF\ \rangle & \approx & q_k\
t^{(on)}_k (0\ , \frac {q^2_k} {2\ \mu_{lj}})\
q_l\ t^{(on)}_l (0\ , \frac {q^2_l} {2\ \mu_{jk}})\\ \nonumber
                      & \approx & \exp {i(\ \delta_k\ +\ \delta_l)}
\sin {\delta_k} \sin {\delta_l}\ \ .
\label{eq:280}
\end{eqnarray}
Here we have used the standard parameterization,
\begin{equation}
t(q, q) = \frac {1} {q(\cot {\delta } - i)}\ \ ,
\label{eq:281}
\end{equation}
with the $\delta$'s being the appropriate s-wave phase shifts. These are
related
to the s-wave scattering length according to,

\begin{eqnarray}
|\langle \ FF\ \rangle| & \approx & \sin \delta_k \sin \delta_l =
\frac {q_k\ a_k\ q_l\ a_l}{(1\ +\ q_k\ a_k)(1\ +\ q_l\ a_l)}\ \ .
\label{eq:283}
\end{eqnarray}
Thus in case of $ q_k\ a_k\ q_l\ a_l \ll 1$ , the factor
$\langle \ FF\ \rangle  \ll 1$ so that
the evaluation of $T^{(el)}_{33}$ to first order should be sufficient.

\begin{center}
\section{Application}
\end{center}
In the present section we apply the procedure described above to analyze
the effects of \fsi on the energy dependence of
$\pi$ and $\eta$ meson production in $NN$ scattering at energies near
their respective production thresholds.
We demonstrate this first for $\pi$ meson production.
For the $pp \to pp \pi^0$ reaction, rescatterings may include
the following sequences :
$pp \to pp\pi^0 \to pp\pi^0$, $pp \to pn\pi^+ \to pp\pi^0$ and
$pp \to d\pi^+ \to pp\pi^0$.
In considering \fsi only the first sequence is included as by definition the
block \fsi includes $elastic$ interactions only. Two-nucleon mechanism
contributions dominate the $pp \to pp \eta$ cross section and any reasonable
calculations of the primary amplitude must include $\pi^+$ production
followed by charge exchange.
The effects of these other channels are assumed to be absorbed
in $M^{(in)}$.
The isosinglet and isotriplet pion-nucleon scattering amplitudes
are calculated from eqn. 20,
using the isosinglet and isotriplet pion nucleon scattering length
values $a_1 = 0.245 fm$ and $a_3 = -0.143 fm$ of ref.\cite{slength}.
The isospin average scattering length of these values is rather small so
that $a\ priori$ the overall $\pi$N interaction effects on the energy
dependence are expected to be very small.
Similarly, the S-wave $pp$ phase
shift is obtained from effective range expansion using the modified
Cini-Fubini-Stanghellini formula\cite{noyes},
\begin{equation}
C^2\ q\ ctn\ \delta + 2\eta_c q h(\eta_c) =
-\frac {1}{a_{pp}} + \frac {1}{2} r_{pp} q^2 -
\frac {p_1\ q^4}{ 1\ +\ p_2 k^2}\ \ ,
\label{eq:281}
\end{equation}
where $a_{pp}$ and $r_{pp}$ denote the $pp$ scattering length and effective
range and,
\begin{equation}
\eta_c = \frac {e^2 E_{lab}} {P_{lab}}\ \ ,
C^2 = \frac {2\pi\ \eta_c} {  (\exp{2\pi\eta_c} - 1)}\ \ ,
h(\eta_c) = \sum_{s=1}^{\infty} \frac {\eta_c^2}{s(s^2 + \eta_c^2)}
- \gamma - \ln {\eta_c}\ \ . \\
\label{eq:281}
\end{equation}
The $p_i$ (i=1,2) are functions of $a_{pp}$ and $r_{pp}$.
In what follows we have used the values
$a_{pp} = -7.82 fm$ and $r_{pp} = 2.7 fm$ of
ref.\cite{noyes}.
To calculate the cross section, the \fsi factor $|Z_{33}|^2$, eqn. 18,
is multiplied by the invariant phase space and then integrated
over the appropriate momenta. The
primary production amplitude is assumed to be constant and is taken outside
the integral.
The energy dependence of the
cross section as obtained with the \fsi factor calculated to first and second
orders are displayed in fig. 4 as a function of $Q_{cm}$, the energy
available in the $cm$ system. The two solutions are practically identical.
The $\langle \ FF\ \rangle $ factor discussed
in the previous
section, is indeed very small so that the sum in eqn. 18 converges and it is
safe to compare data with the predictions corresponding to first order
calculations of $|Z_{33}|^2$.
Comparison with data is made in fig. 5.
The predictions with
the full $FSI$ correction of
eqn. 18 are shown as the solid line.
Those without the $\pi N$ interaction and with only the first disconnected
diagrams of fig. 3b are given by the large-dashed and small-dashed
lines, respectively. The data in the figures are taken from
ref.\cite{meyer}. The curves are arbitrarily normalized to the lowest
energy data point available.
As indicated above the
s-wave $\pi N$ interaction is relatively weak so that the $FSI$ correction
factor is dominated by the strong interaction of the $^1 S_0$ NN system.
Our predictions with the full
\fsi correction factor (solid curve) agrees slightly better with data
compared with those obtained by including the $NN$ force only (dashed line).
Nonetheless, even in this case,
where the overall meson-nucleon interaction is weak,
the effects from disconnected and
completely connected diagrams are comparable.
It is our contention that
the completely connected diagrams of fig. 3b
must be included if one wishes
to interpret the recent precision measurements of $\pi$ production
cross section\cite{meyer,lee}. As demonstrated below
the importance of these completely connected
diagrams becomes more apparent in applying our procedure to $\eta$ production.

For $\eta$ meson production
there are no direct measurements of elastic $\eta N$ scattering and the
information available are extracted indirectly using $\pi N \to \eta N$
data.
Based on $\pi N$ and $\eta N$ coupled channel analysis around the $\eta N$
threshold within an isobar model, Bhalerao and Liu\cite{bahl} suggest a value
$a_{\eta N} = (0.27 + i0.22)\ fm$
for the $\eta N$ scattering length.
With this choice of $a_{\eta N}$ the results
of our analysis with the \fsi factor calculated to first order
are shown in fig. 6. Here as for pion production, the solution with the
\fsi factor calculated to second order is nearly identical.
In contrast with $\pi$ production, this process is
strongly influenced by
the $\eta N$ interaction giving rise to a sharp enhancement of
the cross section very near
threshold and thus reproduces,
rather accurately, the energy dependence of the cross section up to
$Q_{cm} = 20 MeV$. Note also, that in the analysis presented above, the
enhancement factor $|Z_{33}|^2$ is exactly what is required to keep the primary
production amplitude nearly constant over the energy range from
threshold to $Q_{cm} = 20 MeV$.

It is interesting to explore how
sensitive the results are for a different
choice of the $a_{\eta N}$ scattering length.
Since other channels such as the $\pi\pi N$ are also opened,
Wilkin\cite{wilkin} has fixed the imaginary part to be
$\Im [a_{\eta N}] = 0.30 fm$ and  used the $\pi^- p \to \eta p$ data
directly to determine the real part. This procedure has yielded a scattering
length $a_{\eta N} = (0.55\pm0.20 + i0.30)\ fm$. Calculations with
this value are given in fig. 7.
The energy dependence is mostly sensitive to the real part of $a_{\eta N}$
and the results from
our analysis do not support the $central$ value proposed by
Wilkin\cite{wilkin}. It is to be noted also,
that based on partial wave unitarity
relations, an analysis of the $\pi^- p$ backward differential cross section
data yields a value of
$39^o.3  \pm 4^o.7$ for the phase of the $\eta N$
production amplitude\cite{ramesh} a value in agreement with Bhalerao and
Liu\cite{bahl} but not with
Wilkin\cite{wilkin}.

\begin{center}
\section{Summary And Conclusions}
\end{center}

In summary we have developed an approximate
procedure which accounts for final state
effects in the presence of three strongly interacting particles. In analogy
with the Watson's treatment\cite{watson} it was shown that, in the the limit
of extremely strong \fsi the amplitude factorizes into a primary production
amplitude and an elastic scattering amplitude describing a $3 \to 3$
transition. The energy dependence of the cross section in the near vicinity
of the reaction threshold is determined almost entirely by $T^{(el)}_{33}$.
Based on the coherence formalism of Amado\cite{amado}
we argue that the completely connected diagrams corresponding to
two rescatterings may interfere strongly and therefore,
play an important role in determining the energy
dependence of the cross section. In view of the scarcity of knowledge
about the meson-nucleon force, the approximation is particularly useful
near threshold where $T^{(el)}_{33}$
can be estimated using nucleon-nucleon and meson-nucleon scattering data.
Because of the strong NN force the \fsi term dominates.
Detailed analysis of $\pi$ and $\eta$ production
via $NN$ collisions show that the energy dependence of the cross section
can be explained in both cases,
by taking into account the meson-$N$ interaction also.

In the present analysis we have not considered \isi corrections as these vary
slowly with energy and would have little influence on the energy dependence.
For $NN$ scattering at bombarding energies of several 100 MeV and above,
the nuclear and Coulomb distortions are expected to be very small.
The Gamov factor which accounts for Coulomb distortions is equal unity within
2.5$\%$ even for the pion data analyzed above. As previously mentioned \isi
effects can be included
exactly by using distorted waves for the incoming two-proton state.

Finally,
the primary production amplitude
is not treated explicitly so that the cross section scale was arbitrary.
This deficiency can be removed by
incorporating in eqn. 17 a primary production amplitude
from one of the existing perturbative models\cite{germo,lagt,vett}.
Such model calculations were performed for $\eta$ production, taking the
$\eta$N interaction in the primary amplitude to be consistent with that
occurring in the \fsi. They are found to reproduce the energy dependence
as well as the scale of the cross section
and will be published elsewhere\cite{shorer}.

In view
of the large effects due to the $\eta N$ force the analysis presented above
could well be used to study the $\eta N$ force itself for which no direct
experiments are possible.

\newpage
\begin{center}
{\bf Appendix A}
\end{center}

The Fermi-Watson theorem\cite{joachim} can be extended to three-body reactions
by applying unitarity.
Let $T_{23}$ denotes the entire amplitude and let $t^{(el)}_{22'}$
and $T^{(el)}_{3',3}$
denote elastic $2 \to 2$ and $3 \to 3$ amplitudes, respectively.
Then by unitarity,
\begin{equation}
\Im \{ T_{23} \} = - \pi \sum_{2'} t^{(el)}_{22'} \delta (E - E_{2'})
T^{\dagger}_{2'3} - \pi \sum_{3'} T_{23'} \delta (E - E_{3'})
T^{(el) \dagger}_{3'3}\ \ .
\label{eq:10}
\end{equation}
The $\delta$ functions force the two body and three body
elastic scattering amplitudes, $t^{(el)}_{22'}$ and
$T^{(el)}_{3'3}$ to be on mass-shell.

In angular momentum representation the amplitudes of eqn. 24 can be written
in the form,

\begin{equation}
T_{ij}  =  | T_{ij} | \exp {i \delta _{ij}} \ \ .
\label{eq:2}
\end{equation}
Note that $\delta _{2'2}$ stands for two-body phase shift but
$\delta _{3'3}$ is only related to two-body phase shifts of the three
interacting pairs and in itself is not
a measurable quantity.
Then substituting these into the unitarity condition, eqn. 24, leads to the
following constraint on the $\delta$'s,

\begin{eqnarray}
\Im \{ |T_{23}| \exp {i \delta_{23}} \} & = &
- \pi \sum_{2'} |T^{(on)}_{22'}|  |T^{(on)}_{2'3}|
\exp {i (\delta_{22'} - \delta_{2'3})}\nonumber \\
                                        &   &
 - \pi \sum_{3'} |T^{(on)}_{23'}|  |T^{(on)}_{3'3}|
\exp {i (\delta_{23'} - \delta_{3'3})}\ \ .
\label{eq:11}
\end{eqnarray}

In the case of a single channel with the angular momenta of
the final pairs being zero, the phase $\delta_{23}$ from eqn. 26 is,

\begin{eqnarray}
\sin {\delta_{23} } & = &
{ |t^{(on)}_{22}|  \sin {\delta_{22}} -
|T^{(on)}_{33}|  \sin {\delta_{33}} } / \nonumber \\
     &   &
[ |t^{(on)}_{22}|^2
+ |T^{(on)}_{33}|^2
- 2 |t^{(on)}_{22}|  |T^{(on)}_{33}|
\cos {(\delta_{33}  - \delta_{22})} ]^{1/2}\ \ .
\label{eq:12}
\end{eqnarray}

Thus the phases of the
rescattering amplitudes and their norms, determine
the overall phase of the $2 \to 3$ transition amplitude,
just as the case is
for a two body reactions.
In the limit of very weak \isi (vanishing $t^{(el)}_{22}$),

\begin{equation}
\sin {\delta_{23} } \approx  \sin { \delta_{33}} \ \ ,
\label{eq:5}
\end{equation}
so that the overall phase of $T^{(el)}_{23}$
becomes identical to that of the elastic
scattering amplitude $T^{(el)}_{33}$.
Furthermore, the amplitudes and phases of the strong rescatterings in the
entrance and exit channels are
related through eqn. 26 by,

\begin{equation}
|t^{(el)}_{22}|\  ( \sin {\delta_{22} } + \pi \ |t^{(el)}_{22}| ) =
|T^{(el)}_{33}|\  ( \sin {\delta_{33} } + \pi \ |T^{(el)}_{33}| ) \ \ .
\label{eq:5}
\end{equation}

For a real production process
the amplitudes $t^{(el)}_{22}$ and $T^{(el)}_{33}$ must
be independent so that the lhs and rhs of eqn. 27 must be equal to a
constant. In the limit of very weak \isi this constant must vanish so that,
\begin{equation}
|T^{(el)}_{33}| = \sin {\delta_{33} }/ \pi\ \ .
\label{eq:7}
\end{equation}
Formally, this is identical in form to the expression
assumed by Watson\cite{watson} for two-body processes.

\begin{center}
{\bf Appendix B}
\end{center}

We evaluate the contribution of diagrams with two rescatterings.
As an example consider the diagram shown in fig. 8 (crossed lines denote on
mass-shell states).
In the CM system,
in each of the three-body states of the diagram, there are only two
independent momenta which are taken to be ${\bf p}_j$ the momentum of particle
$j$ and ${\bf q}_j$ the relative momentum of the remaining $l-k$ pair.
Adopting the convention
that $(j,l,k)$ are always cyclic, ${\bf q}_j$ is defined as
\begin{equation}
{\bf q}_j  = \frac {  (m_k {\bf p}_l - m_l {\bf p}_k )}
{( m_l + m_k)}\ \ .
\label{eq:2.39}
\end{equation}
In terms of these momenta the total kinetic energy is ,
\begin{equation}
H_{0}  =  \frac {p^2_j}{2 \mu_j} + \frac {q^2_j}{2 \mu_{lk}}\ \ .
\label{eq:2.39}
\end{equation}
Here $\mu_{lk}$ is the reduced mass of the $l$ and $k$ particles and $\mu_j$
is the reduced mass of the $j$ particle, \ie,
\begin{eqnarray}
\mu_{lk}  =  \frac {m_l m_k}{( m_l + m_k )}\ \ ;\
\mu_j = \frac {m_j (m_l + m_k) } {(m_j + m_l + m_k)} \ \ .
\label{eq:2.39}
\end{eqnarray}
Now the contribution of the diagram, fig. 8, can be written in the form,
\begin{eqnarray}
C_{31} & = & \int d{\bf p}_3' d{\bf q}_3'
\delta (E - \frac {p_3'^2}{2 \mu_3}- \frac {q_3'^2}{2 \mu_{12} })
M^{(in)}({\bf p}_3',{\bf q}_3')\\ \nonumber
       &   & \int d{\bf p}_1''d{\bf q}_1'' \delta ({\bf p}_3' - {\bf p}_3'')
t_3 ({\bf q}_3', {\bf q}_3''; E^+ - \frac {p_3''^2}{2 \mu_3})\\ \nonumber
       &   &  [E^+ - \frac {p_1''^2}{2 \mu_1} -
\frac {q_1''^2} {2 \mu_{23}}]^{-1}
\delta ({\bf p}_1'' - {\bf p}_1) )
t_1 ({\bf q}_1'', {\bf q}_1; E^+ - \frac {p_1^2}{2 \mu_1})\ \ .
\label{eq:2.39}
\end{eqnarray}
The integration over ${\bf p}_1''$ and ${\bf p}_3'$ are immediate and we
obtain,
\begin{eqnarray}
C_{31} & = & \int d{\bf q}_3'
\delta (E - \frac {p_3''^2}{2 \mu_3}- \frac {q_3'^2}{2 \mu_{12} })
M^{(in)}({\bf p}_3'',{\bf q}_3')\\ \nonumber
       &   & \int d{\bf q}_1''
t_3 ({\bf q}_3', {\bf q}_3''; E^+ - \frac {p_3''^2}{2 \mu_3})\\ \nonumber
       &   &  [E^+ - \frac {p_1^2}{2 \mu_1} -
\frac {q_1''^2} {2 \mu_{23}}]^{-1}
t_1 ({\bf q}_1'', {\bf q}_1; E^+ - \frac {p_1^2}{2 \mu_1})\ \ .
\label{eq:2.39}
\end{eqnarray}
Here $t_3$ is a half off mass-shell two-body matrix element
with only the ${\bf q}_3''$ being off-shell momentum.
(From a view point of invariant perturbation theory only one leg of
$t_j$ is off mass-shell so that it behaves like a particle form-factor).
Expanding $t_j$ in partial waves and introducing the
Kowalski-Noyes half-shell function\cite{kowalski}
$f_j(\lambda ,{\bf q}_j,{\bf q}'')$
allows writing these for a partial wave $\lambda$ as,
\begin{equation}
t_{j}(\lambda ,{\bf q}_j,{\bf q}_j'';E^+)  =
t^{(on)}_j (\lambda ,q^2_j /2 \mu_j)
\ f_j (\lambda ,{\bf q}_j,{\bf q}_j'')\ \ .
\label{eq:2.39}
\end{equation}
Then for slowly varying $M^{(in)}$ and at energies near threshold one obtains,
\begin{equation}
C_{31}  =  M^{(in)} (0,0)\  q_3\  t^{(on)}_3 (0, \frac {q_3^2 }{2 \mu_{12}})
\ t^{(on)}_1 (0, \frac {q_1^2 }{ 2 \mu_{23}})\  I_{31}\ \ ,
\label{eq:2.39}
\end{equation}
where $I_{31}$ is an integral over the
Kowalski-Noyes half-shell functions,
\begin{equation}
I_{31}  =  \frac {1} {(2\pi)^2 \ \mu_{23} \ q_3} \int d{\bf q}_1''
f_3 (0,{\bf q}_3,{\bf q}_3'')
[ E - \frac {p^2_1}{2 \mu_1} - \frac {q''^2_1}{2 \mu_{23}}]^{-1}
f_1 (0,{\bf q}_1,{\bf q}_1'')\ \ ,
\label{eq:2.39}
\end{equation}
where,
\begin{equation}
{\bf q}_3'' = -\frac {m_1}{m_1 + m_2 } {\bf q}_1'' -
\frac { m_2 (m_1 + m_2 + m_3) } {(m_1 + m_2) (m_1 + m_3)} {\bf p}_1\ \ .
\label{eq:2.39}
\end{equation}
If we neglect off mass-shell effects the integrals $I_{jl}$ reduce to unity.

\newpage
\begin{center}
{\bf Appendix C}
\end{center}

We consider here second order contributions to the transition amplitude. As
indicated in the text these are obtained by multiplying first order
contributions with a factor (FF). We attempt
demonstrating this for the diagram shown in Fig. 9. This can be written as,
\begin{eqnarray}
D_{313} & = & \int d{\bf p}_3'\ d{\bf q}_3'\ d{\bf p}_1''\ d{\bf q}_1''\
d{\bf p}_3'''\ d{\bf q}_3'''\
\delta (E - \frac {p_3'^2}{2 \mu_3}- \frac {q_3'^2}{2 \mu_{12} })\\ \nonumber
       &   & M^{(in)}({\bf p}_3',{\bf q}_3')\
t_3 ({\bf q}_3', {\bf q}_3''; E^+ - \frac {p_3'^2}{2 \mu_3})\
\delta ({\bf p}_3' - {\bf p}_3'')\
\\ \nonumber
       &   &  [E^+ - \frac {p_1''^2}{2 \mu_1} -
\frac {q_1''^2} {2 \mu_{23}}]^{-1} \
\delta ({\bf p}_1'' - {\bf p}_1''') \
t_1 ({\bf q}_1'', {\bf q}_1'''; E^+ - \frac {p_1'''^2}{2 \mu_1})\\ \nonumber
       &   &  [E^+ - \frac {p_3'''^2}{2 \mu_3} -
\frac {q_3'''^2} {2 \mu_{12}}]^{-1} \
\delta ({\bf p}_3''' - {\bf p}_3) \
t_3 ({\bf q}_3''', {\bf q}_3; E^+ - \frac {p_3^2}{2 \mu_3})\ \ .
\label{eq:2.40}
\end{eqnarray}
Integrating over ${\bf p}_3'$, ${\bf p}_1''$ and ${\bf p}_3'''$ and arranging
terms leads to,
\begin{eqnarray}
D_{313} & = & {\Huge \{ }\  q_1\
t^{(on)}_1 (0 \ ,\  \frac {q_1'''^2}{2 \mu_{23}})\
q_3\  t^{(on)}_3 (0\ ,\  \frac {q_3^2}{2 \mu_{12}})\ \frac {1} { q_1\ q_3}
\ \int  d{\bf q}_1''
\frac {1} { [E^+ - \frac {p_1'''^2}{2 \mu_1} - \frac {q_1''^2} {2 \mu_{23}}]}
\\ \nonumber
       &   &  \int  d{\bf q}_3''' \frac {f_1(0,{\bf q}_1'',{\bf q}_1''')\
f_1(0,{\bf q}_3''',{\bf q}_3)}
{[E^+ - \frac {p_3^2}{2 \mu_1} - \frac {q_3'''^2} {2 \mu_{12}}]}\
{\Large \}} \\ \nonumber
        &   & \int d{\bf q}_3' \ M^{(in)}({\bf p}_3',{\bf q}_3')\
\delta (E - \frac {p_3''^2}{2 \mu_3}- \frac {q_3'^2}{2 \mu_{12} })
t_3 (0, \frac {q_3'^2}{2 \mu_{12}})
f_3(0, {\bf q}_3', {\bf q}_3'')\ \ .
\label{eq:241}
\end{eqnarray}
We now notice that the last integral is the first order contribution from
the disconnected diagram (9.a) with a single scattering
block between particles 1 and 2 while the
expression in the curly bracket is just the (FF) factor.
An order of magnitude of this
factor can be obtained using on mass-shell values for
the Kowalski-Noyes functions. This leads to,
\begin{equation}
\langle \ FF\ \rangle \approx  q_1\
t^{(on)}_1 (0\ ,\  \frac {q_1'''^2}{2 \mu_{23}})\
q_3\  t^{(on)}_3 (0\ ,\  \frac {q_3^2}{2 \mu_{12}})\  \ .
\label{eq:242}
\end{equation}
Similar expressions can be calculated for the completely connected diagrams
(3.b) as well.

\vspace{0.5cm}

{\bf Acknowledgments}
This work was supported in part
by Israel Ministry Of Science and Technology and the Israel Ministry Of
Absorption. One of us ( A. M. ) thanks C. Wilkin for suggestions
and stimulating discussions concerning
the results reported in this work.

\vspace{0.4cm}

\begin{figure}
\vspace{6.0in}
\includegraphics{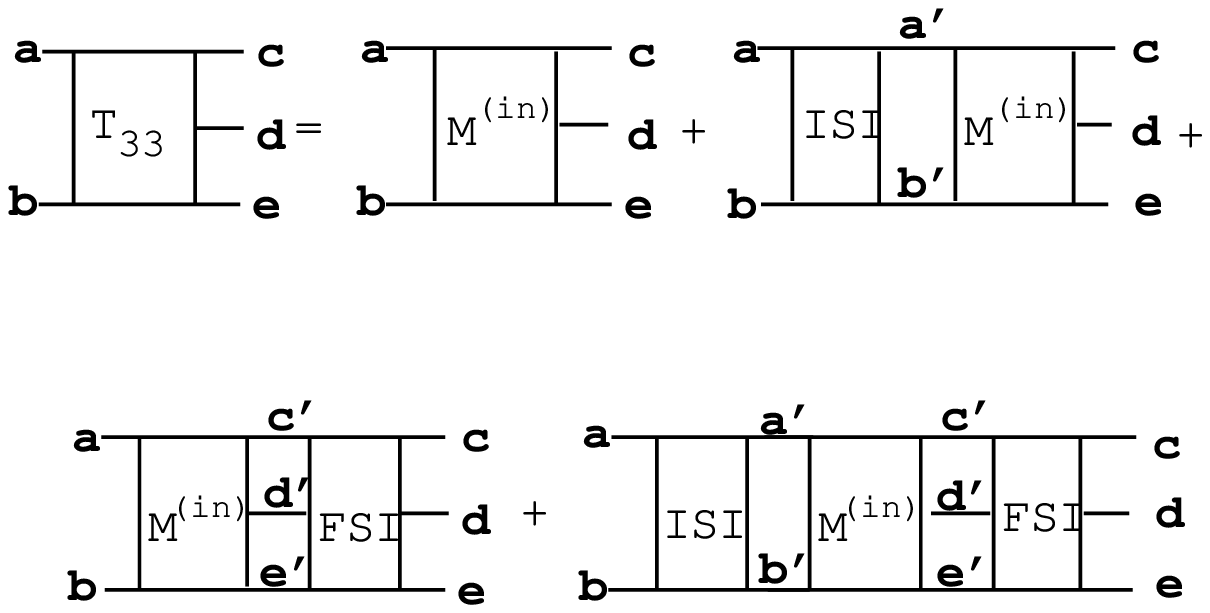}
\vskip 0.5 in
\caption{  Graphical representation of the transition amplitude.
The block $M^{(in)}$ represents the primary production
mechanism of the reaction and contains all possible inelastic transitions from
a two-particle state to a three-particle state (see text).
The \isi and  $FSI$ blocks represent elastic rescattering in the entrance
and exit channels, respectively.
}
\label{mechanism}
\end{figure}

\begin{figure}
\vspace{4.5in}
\includegraphics{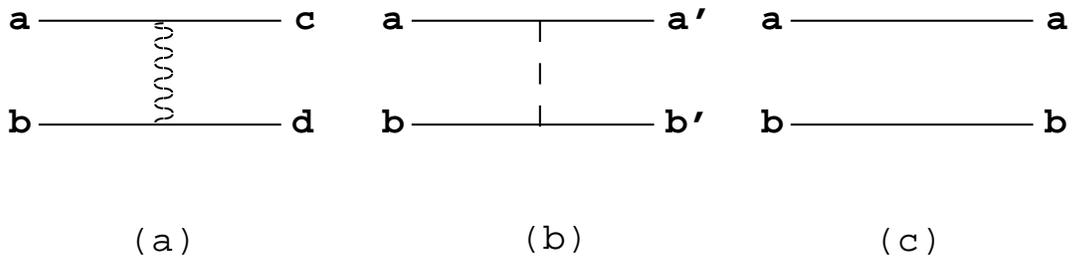}
\vskip 0.5 in
\caption{ Elementary inelastic(2a) and elastic(2b) interactions. A
wavy line represents the primary interaction, such that if it were zero
the reaction $ab \to cd$ would not occur. A dotted line represents an elastic
interaction. The disconnected diagram(2c)
describes non-interacting particles.
}
\label{interactions}
\end{figure}

\begin{figure}[t]
\vspace{7.5in}
\includegraphics{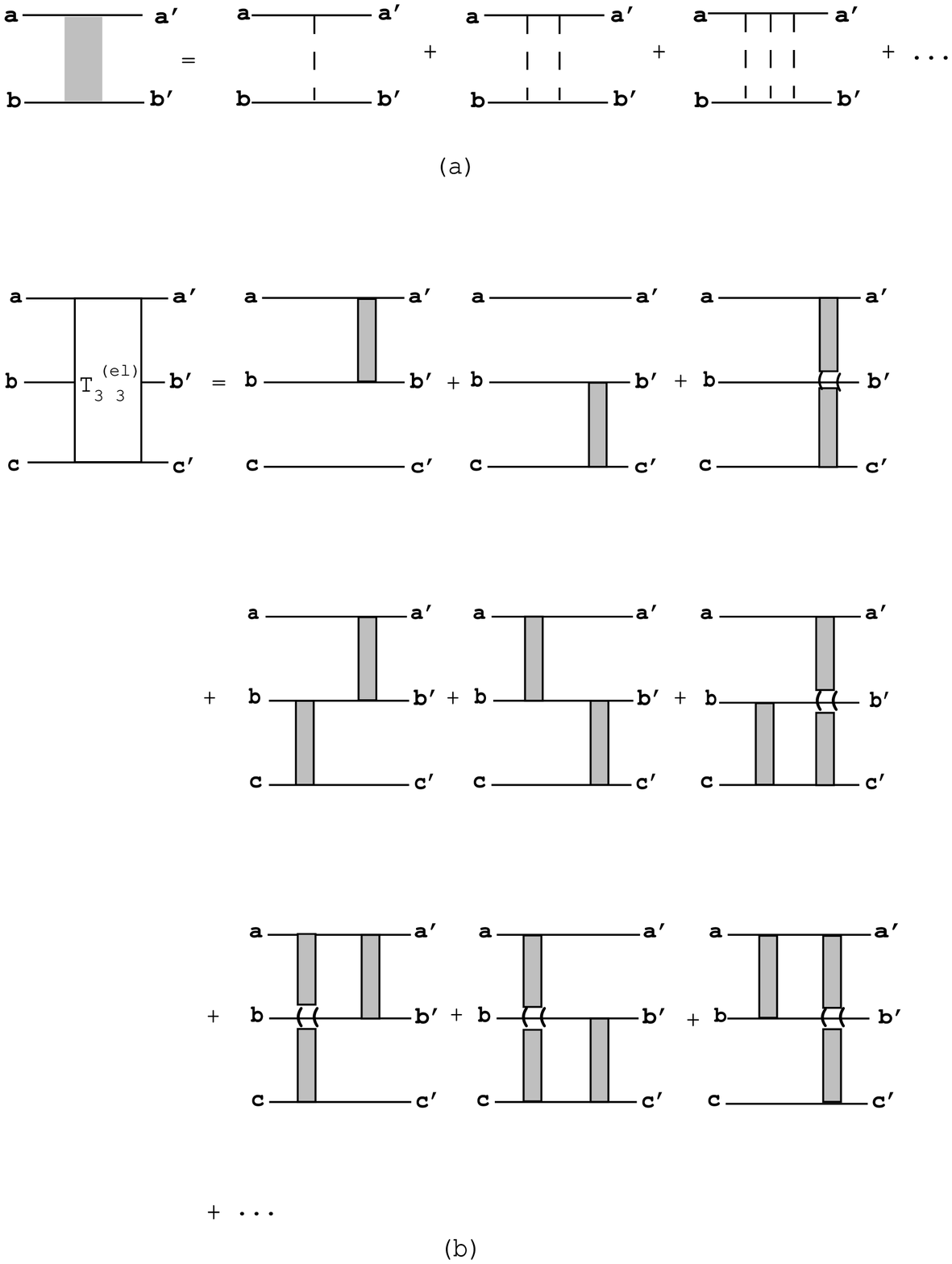}
\vskip 0.5 in
 \caption{ Graphical representation of elastic amplitudes: a) the
amplitude $t^{(el)}_{22}$ for $ab \to ab$, b) the amplitude $T^{(el)}_{33}$
for $cde \to cde$; the blocks represent two-body scattering amplitudes in
three-body space similar to the one defined in 3a.
}
    \label{amplitudes}
\end{figure}

\begin{figure}[t]
\vspace{7.0in}
\includegraphics{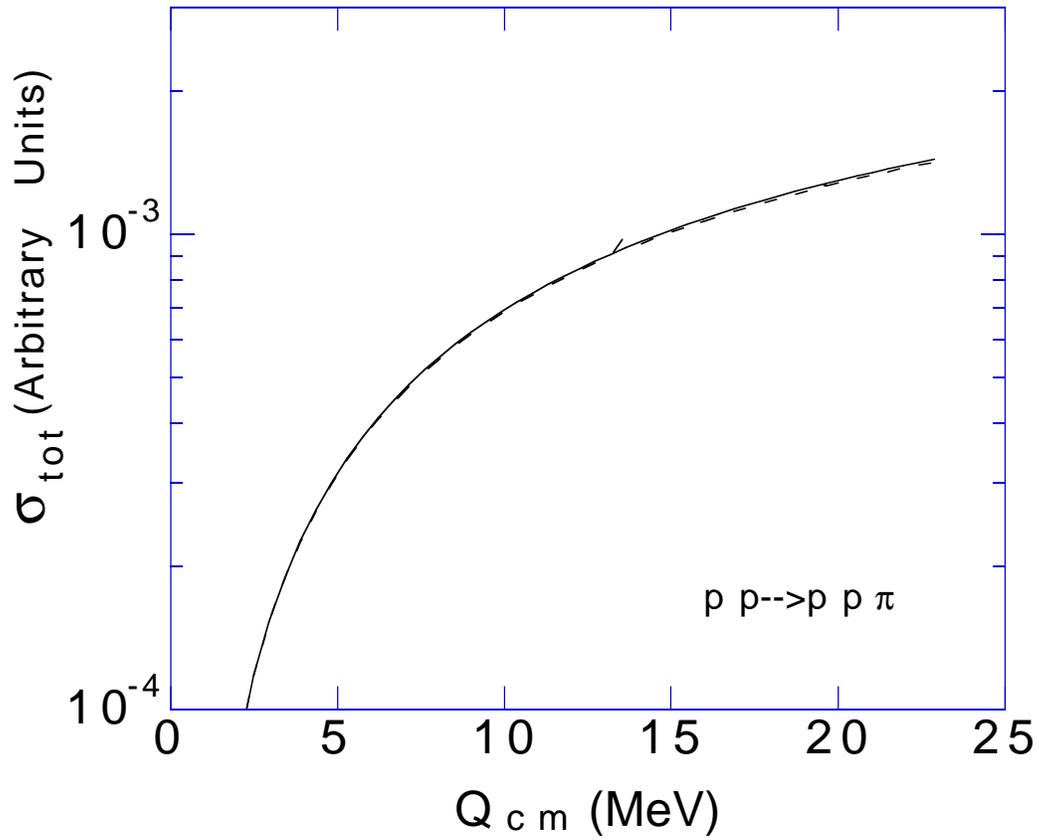}
\vskip 0.5 in
 \caption{The $pp \to pp \pi^0$ cross section $vs.$
the energy available in the $cm$ system, $Q_{cm}$.
The solid line represents predictions with the $FSI$ correction factor
calculated to first ($T_{33}^{(el)} \approx [ t + F t ]$). The dashed line
is that obtained with the \fsi taken to second order,$i.e.$,
$T_{33}^{(el)} \approx (1 + FF) [ t + F t ]$ (see text).
}
\label{pppion}
\end{figure}

\begin{figure}[t]
\vspace{7.0in}
\includegraphics{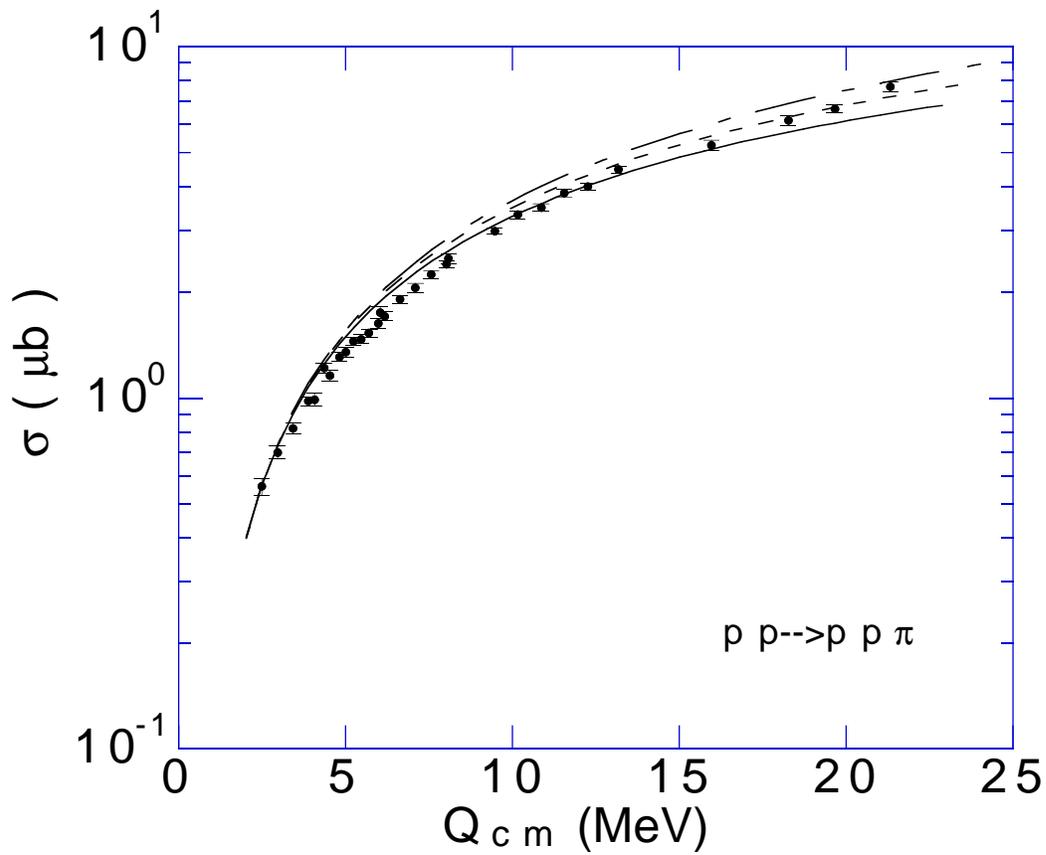}
\vskip 0.5 in
 \caption{The $\pi$ meson production
cross section $vs.$ the energy available in
the $cm$ system, $Q_{cm}$.
The predictions with the full $FSI$ correction factor of eqns. 20 are shown
as the solid line. Those without the meson-nucleon interaction and with the
disconnected diagrams of fig. 3b are given by the large-dashed
and small-dashed lines, respectively.
All curves are normalized to the lowest data point.
}
\label{pppion}
\end{figure}

\begin{figure}[t]
\vspace{5.0in}
\includegraphics{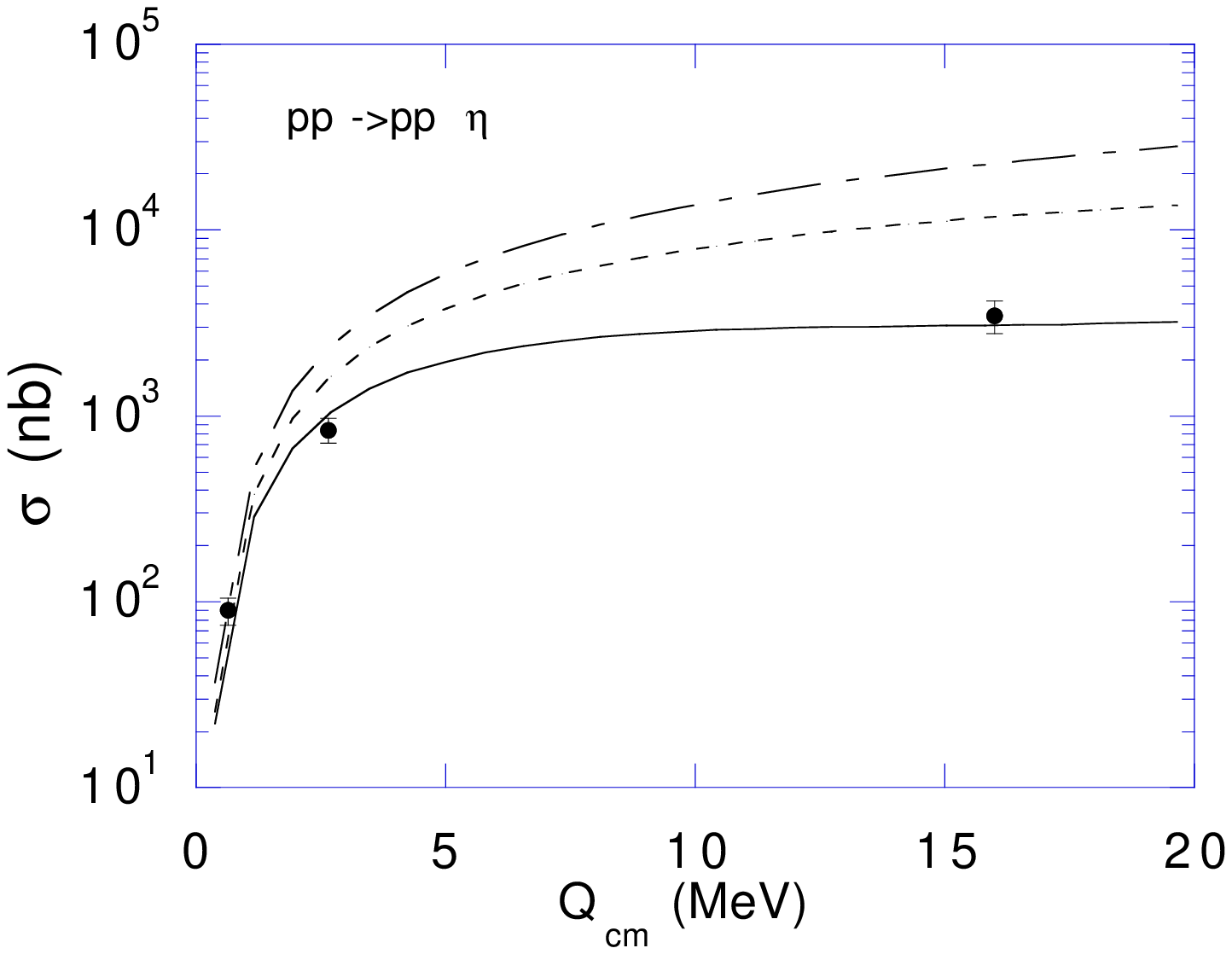}
\vskip 0.5 in
\caption{ The $\eta$ meson production cross section $vs.$ $Q_{cm}$.
The $\eta N$ scattering length is taken from ref. [18].
(see captions of fig. 5)  }
\label{ppetabal}
\end{figure}

\begin{figure}[t]
\vspace{6.0in}
\includegraphics{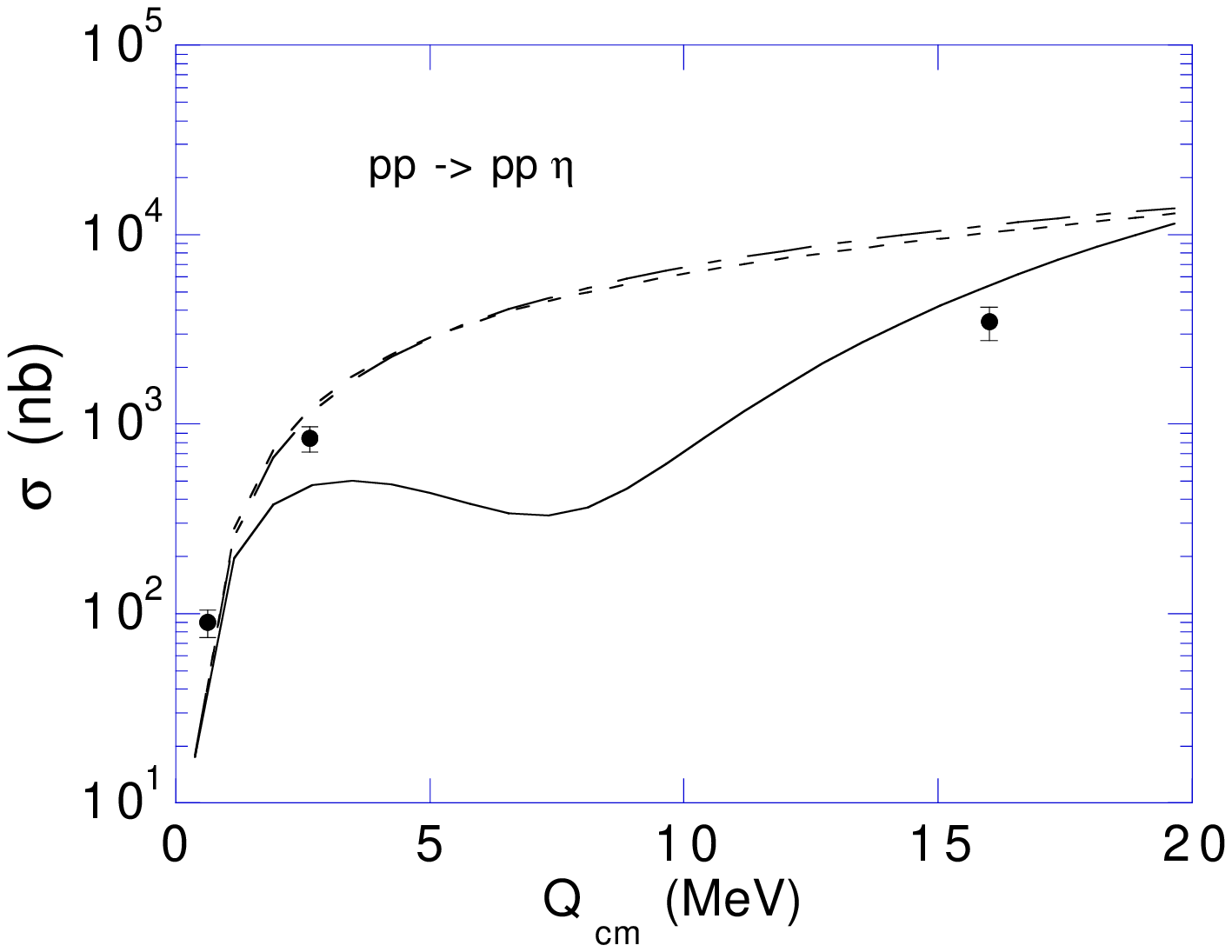}
\vskip 0.5 in
\caption{ The $\eta$ meson production cross section $vs.$ $Q_{cm}$.
The $\eta N$ scattering length is taken from
ref. [3]
(see captions of fig. 5)
}
    \label{msat}
\end{figure}

\begin{figure}[t]
\vspace{6.5in}
\includegraphics{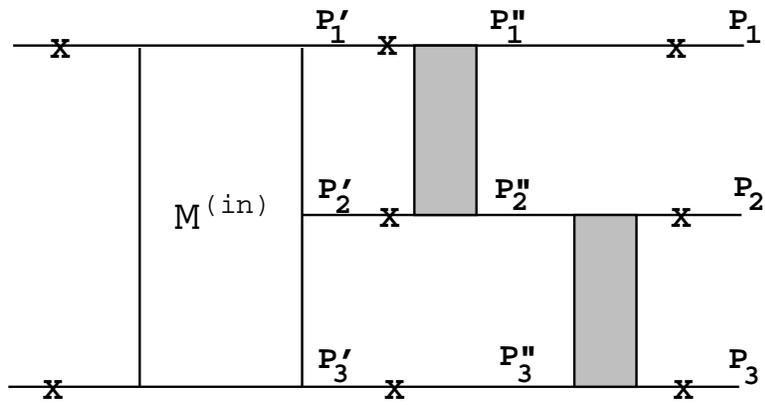}
\vskip 0.5 in
 \caption{ Typical completely connected diagram with two subsequent
rescatterings. Crossed lines denote on mass-shell states.
}
    \label{diagram}
\end{figure}

\begin{figure}[t]
\vspace{6.5in}
\includegraphics{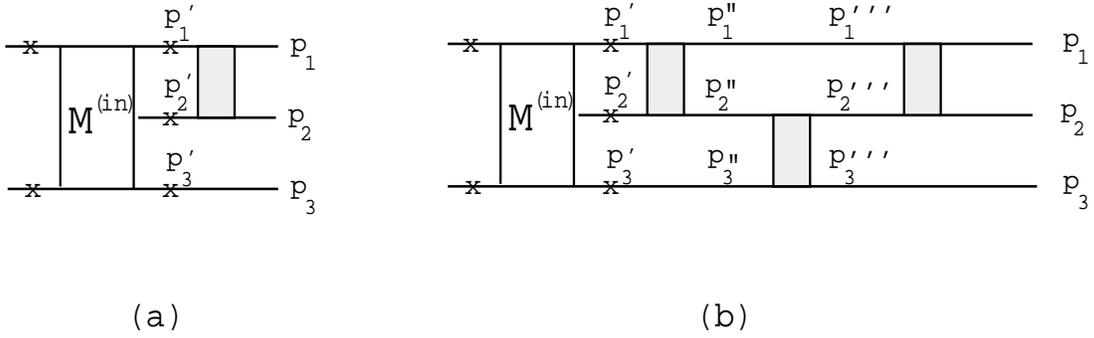}
\vskip 0.5 in
 \caption{ An example of diagrams contributing to the $T^{(el)}_{33}$ amplitude
to second order : (a) a disconnected diagram which contributes to first order,
(b) a completely connected diagram obtained from diagram (a) with the FF factor
(see text).
}
    \label{diagrams}
\end{figure}

\end{document}